\documentclass[sigconf]{acmart}

\usepackage{tikz}
\graphicspath{ {./Figures/} }

\newcommand{\A}{{A}}
\newcommand{\bit}{\{0,1\}}
\newcommand{\ignore}[1]{}
\newcommand{\PFL}{{ \Pi_{\sf PPFL}}}
\newcommand{\secparam}{{\lambda}}
\newcommand{\Setup}{{\sf Setup}}

\newcommand{\abs}[1]{\left|{#1}\right|}
\newtheorem{myprot}{Protocol}

\usepackage{tabularx}
\newcounter{protocol}
\newenvironment{protocol}[1]
  {\par\addvspace{\topsep}
   \noindent
   \tabularx{\linewidth}{@{} X @{}}
    \\\hline
    \refstepcounter{protocol}\textbf{Protocol \theprotocol} #1 
    \\\hline}
  {\hrule
  }
\usepackage{enumitem}

\copyrightyear{2020} 
\acmYear{2020} 
\setcopyright{acmcopyright}\acmConference[ICAIF '20]{ACM International Conference on AI in Finance}{October 15--16, 2020}{New York, NY, USA}
\acmBooktitle{ACM International Conference on AI in Finance (ICAIF '20), October 15--16, 2020, New York, NY, USA}
\acmPrice{15.00}
\acmDOI{10.1145/3383455.3422562}
\acmISBN{978-1-4503-7584-9/20/10}

\begin{document}

\title{Differentially Private Secure Multi-Party Computation for Federated Learning in Financial Applications}

\author{David Byrd}
\email{db@gatech.edu}
\affiliation{
    \institution{School of Interactive Computing \\
                 Georgia Institute of Technology}
    \city{Atlanta}
    \state{Georgia}
    \postcode{30308}
}

\author{Antigoni Polychroniadou}
\email{antigoni.poly@jpmorgan.com}
\affiliation{
    \institution{J.P. Morgan AI Research}
    \city{New York}
    \state{New York}
    \postcode{10017}
}

\begin{abstract}
Federated Learning enables a population of clients, working with a trusted server, to collaboratively learn a shared machine learning model while keeping each client's data within its own local systems.
This reduces the risk of exposing sensitive data, but it is still possible to reverse engineer information about a client's private data set from communicated model parameters.
Most federated learning systems therefore use differential privacy to introduce noise to the parameters.  This adds uncertainty to any attempt to reveal private client data, but also reduces the accuracy of the shared model, limiting the useful scale of privacy-preserving noise.
A system can further reduce the coordinating server's ability to recover private client information, without additional accuracy loss, by also including secure multiparty computation.
An approach combining both techniques is especially relevant to financial firms as it allows new possibilities for collaborative learning without exposing sensitive client data.  This could produce more accurate models for important tasks like optimal trade execution, credit origination, or fraud detection.
The key contributions of this paper are: We present a privacy-preserving federated learning protocol to a non-specialist audience, demonstrate it using logistic regression on a real-world credit card fraud data set, and evaluate it using an open-source simulation platform which we have adapted for the development of federated learning systems.
\end{abstract}


\maketitle

\section{Introduction}

Modern financial firms routinely need to conduct analysis of large data sets stored across multiple servers or devices.  A typical response is to combine those data sets into a single central database, but this approach introduces a number of privacy challenges: The institution may not have appropriate authority or permission to transfer locally stored information, the owner of the data may not \emph{want} it shared, and centralization of the data may worsen the potential consequences of a data breach.

For example, the mobile app ai.type collected personal data from its users' phones and uploaded this information to a central database.  Security researchers gained access to the database and obtained the names, email addresses, passwords, and other sensitive information of 31 million users of the Android version of the app. Such incidents highlight the risks and challenges associated with centralized data solutions. \cite{aitypeArticle}

In this section, we motivate our approach while providing an extensive non-technical overview of the underlying techniques.

\subsection{Federated Learning}
One approach to mitigate the mentioned privacy concerns is to analyze the multiple data sets separately and share only the resulting insights from each analysis.  This approach is realized in a recently-introduced technique called federated analysis.  \cite{bonawitz2017practical}  Federated \emph{learning}, already adopted by large companies like Google, allows users to share insights (perhaps the parameters of a trained model) from the data on their laptops or mobile devices \emph{without} ever sharing the data itself, typically as follows:

\begin{enumerate}
    \item Users train a local model on their individual data.
    \item Each user sends their model weights to a trusted server.
    \item The server computes an average-weight shared model.
    \item The shared model is returned to all of the users.
    \item Users retrain a local model starting from the shared model.
\end{enumerate}

For instance, email providers could use federated learning to reduce the amount of spam their customers receive.  Instead of each provider using its own spam filter trained from its customers' reported spam email, the providers could combine their models to create a shared spam-detection mechanism, without sharing their individual customers' reported spam emails.  For a survey of recent advances in federated learning, see Kairouz et al.  \cite{kairouz2019advances}

It is still possible, however, for a malicious party to potentially compromise the privacy of the individual users by inferring details of a training data set from the trained model's weights or parameters ~\cite{shokri2017membership,nasr2018comprehensive}. It is important to protect sensitive user information while still providing highly accurate inferences.

\subsection{Differential Privacy}
Simply anonymizing data is no longer enough to guarantee the privacy of individuals whose information has been collected, due to the increasing prevalence of database reconstruction attacks and re-identification from correlated data sets.

Differential privacy  \cite{dwork2006our,mcsherry2007mechanism} can help prevent such reverse engineering by adding noise to the input data set, to intermediate calculations, or to the outputs.  For example, in Step 2 of the Federated Learning process above, each client can add randomly-generated values to its model weights before transmission.  Then, even if the data is reverse engineered, it is \emph{not} the exact data of any user.

More formally, differential privacy is a mathematical concept that guarantees statistical indistinguishability for individual inputs by perturbing values.  The use of differentially-private machine learning algorithms in centralized settings is widely discussed in the literature, and the technique has been adopted by major companies; for example, Apple uses it in web search auto-completion.  The application of differential privacy adds a layer of randomness so that adversaries with additional information still have uncertainty over the original value.  There is an obvious trade-off: adding randomness to the collected data preserves user privacy at the cost of accuracy.  Proper application of differential privacy ensures that meaningful insights can still be derived from the aggregated data.

\subsection{Secure Multi-Party Computation}
Achieving a desired level of differential privacy can require adding a great deal of accuracy-reducing noise into the mix.  An alternative method which guarantees privacy \emph{without} compromising accuracy is secure multi-party computation (MPC) \cite{GMW87}.  Using MPC, multiple parties collaborate to compute a common function of interest without revealing their private inputs to other parties.  An MPC protocol is considered \emph{secure} if the parties learn only the final result, and no other information.

For example, a group of employees might want to compute their average salary without any employee revealing their individual salary to any other employee.  This task can be completed using MPC, such that the only information revealed is the result of the computation (i.e. the average salary).  If each pair of employees holds a large, arbitrary, shared number, such that one employee will add it to their salary and the other will subtract it, then the result of the computation will not change, but no one will know any employee's real salary.

The same idea can be applied to federated learning by having the parties use a secure weighted average protocol, under which each client encrypts their model weights, but the server can still calculate the weighted average on the encrypted data.

\subsection{Secure Federated Learning}
Considering all of the above, we arrive at the idea of secure federated learning, in which clients encrypt the model weights sent in Step 2 of Federated Learning.  Assuming the encryption scheme is chosen appropriately, the server will still be able to perform the necessary calculation on the encrypted data, but will not be able to discover the original weights for any user.  A recent line of investigation has constructed secure federated learning using techniques from MPC.  \cite{bonawitz2017practical, jayaraman2018distributed}

MPC protects the computation inputs from exposure to the server, but the exact final result is revealed to all parties by design.  Unfortunately, for some types of computation, the final result can be used to reveal information about the inputs.  For example, in the case of employees computing their average salary, once the result is known, if all but one of the employees work together, they can easily determine the salary of the final employee given the output (average salary).  A secure learning approach based only on MPC may not be ideal for these cases.

By applying differential privacy on top of MPC, we can construct a federated learning system that protects from even this type of extreme collusion attack.  If each client adds noise to its model weights before sending, the final calculation will still be accurate within known bounds, but we will eliminate the possible leakage of any inputs from the output.  In a solution which used \emph{only} differential privacy, the server would know the ``noisy'' private weights of each user.  In the solution which combines MPC and differential privacy, the noisy weights sent to the server are also encrypted such that the server can calculate the result, but cannot infer anything about even the noisy weights of any particular user.  The system is thus now fully private.

\subsection{Differentially Private Secure Multi-Party Computation for Federated Learning} A protocol such as the one presented here, which combines federated learning, differential privacy, and secure multi-party computation, should be of particular interest in the finance space.  These firms operate under substantial regulation with respect to the use, protection, and disclosure of client information.  Data sharing, even within a firm, is thus often difficult to achieve, with negative impacts in the ability to harness new techniques in artificial intelligence (AI) to improve key performance indicators at the firm, such as accurate estimation of loan failure rates, reduction of financial market transaction costs, or optimization of product pricing.

This combination approach can improve internal data protections while still enabling the application of powerful AI to the company's data.  Now each client, server, or device's data can be kept securely in its originating silo, where local model training can safely occur, and the trained models can be shared and combined in an encrypted and differentially private manner.  The data silos can thus each contribute to the overall organization learning an accurate, useful, and directly applicable model without increasing the exposure risk of any client's data.  And while no firm wants to give away a competitive advantage, the protocol can also improve models through secure inter-firm collaboration to lower market execution costs or more accurately price the risk component of a loan product, benefiting all participants.

We demonstrate our approach to differentially private secure multi-party aggregation for federated learning by application to a well-known credit card fraud data set \cite{dal2015calibrating}, and show that client populations of varying size can collaboratively build a fraud detection model without sharing or revealing their local data.  We note that Jayaraman et al \cite{jayaraman2018distributed} approached the same problem with heavier cryptography tools, on a data set that is not finance related, and without an end-to-end simulation of the protocol.

The key contributions of this paper are to introduce the finance community to recent advances in secure federated learning, to provide a complete open-source platform on which such protocols can be developed, and to demonstrate a protocol that enables secure learning of a shared fraud detection model in at most 30 protocol iterations on an extremely class-imbalanced real world data set.

\section{Background and Related Work}

In this section, we provide a formal description of the important components underlying our approach.

\subsection{Secure Multiparty Computation}
\label{sec:overview-MPC}

Consider $n$ parties $P_1,\ldots,P_n$ that hold private inputs $x_1 ,\ldots ,x_n$ and wish to compute some arbitrary function $(y_1,\ldots,y_n) = f(x_1,\ldots,x_n)$, where the output of $P_i$ is $y_i$. Secure Multi-Party Computation (MPC) enables the parties to compute the function using an interactive protocol, where each party $P_i$ learns exactly $y_i$, and nothing else. Seminal results established in the 1980s \cite{GMW87} show that \emph{any} computation can be emulated by a secure protocol. 

It is important that the security of the protocol be preserved even in the presence of adversarial behavior.  For example, several leading banks might collaborate to learn an improved model to minimize the transaction costs associated with fulfilling client orders in a financial market.  The privacy of each honest bank's individual client orders should be preserved even if other banks collude by pooling their information, revealing their encryption offsets, or deviating from the specified protocol.

In this work, we focus on a semi-honest adversary who follows the protocol specification, but may attempt to learn honest parties' private information from the messages it receives, or to collude with other parties to learn private information.  This is the same level of security contemplated in prior works in our setting.

\subsection{Differential Privacy}

Differential privacy states that if there are two databases that differ by only one element, they are statistically indistinguishable from each other. In particular, if an observer cannot tell whether the element is in the dataset or not, she will not be able to determine anything else about the element either. 

\begin{definition}
    ($\epsilon$-differential privacy  \cite{dwork2006calibrating}) For any two neighboring datasets $\mathcal{D}_1\sim\mathcal{D}_2$ that differ by one element, a randomized mechanism $\mathcal{A}$: $\mathcal{D}\rightarrow \mathcal{O}$ preserves $\epsilon$-differential privacy ($\epsilon$-DP) when there exists $\epsilon > 0$ such that,
    \begin{equation}
    \text{Pr [}\mathcal{A}(D_1)\in \mathcal{T} \text{]}\leq e^\epsilon \text{ Pr [}\mathcal{A}(D_2)\in \mathcal{T}\text{]}    
    \end{equation}
    holds for every subset $\mathcal{T} \subseteq \mathcal{O}$, where $\mathcal{D}$ is a dataset, $\mathcal{T}$ is the response set, and $\mathcal{O}$ depicts the set of all outcomes.
\end{definition}

The value $\epsilon$ is used to determine how strict the privacy is. A smaller  $\epsilon$ gives better privacy but worse accuracy. Depending on the application $\epsilon$ should be chosen to strike a balance between accuracy and privacy. 

\begin{definition}
    (Global Sensitivity \cite{dwork2006calibrating}) For a real-valued query function $q: \mathcal{D} \rightarrow \mathbb{R}$, where $\mathcal{D}$ denotes the set of all possible datasets, the global sensitivity of $q$, denoted by $\Delta$, is defined as
    \begin{equation}
        \Delta = \max_{\mathcal{D}_1\sim\mathcal{D}_2} |q(\mathcal{D}_1)-q(\mathcal{D}_2)|,
    \end{equation}
    for all $\mathcal{D}_1\in\mathcal{D}$ and $\mathcal{D}_2\in\mathcal{D}$ .
\end{definition}

The sensitivity is defined as the maximum effect of any single input of the function on the output, and should be concealed to preserve privacy.

\subsubsection*{Laplacian Mechanism}
One of the most well-known techniques in differential privacy is the Laplacian mechanism, which uses random noise $X$ drawn from the symmetric Laplacian distribution.  The zero-mean Laplacian distribution has a symmetric probability density function $f(x)$ with a scale parameter $\lambda$ defined as: 
\begin{equation}
    f(x) = \frac{1}{2\lambda}e^{-\frac{|x|}{\lambda}}.
\end{equation}
Given the global sensitivity, $\Delta$, of the query function $q$, and the privacy parameter $\epsilon$, the \textit{Laplacian mechanism} $\mathcal{A}$ uses random noise $X$ drawn from the Laplacian distribution with scale $\lambda = \frac{\Delta}{\epsilon}$. The Laplacian mechanism preserves $\epsilon$-differential privacy \cite{dwork2006our}.

\subsection{Training Local Logistic Regression Classifiers}

Logistic regression is a machine learning algorithm used to solve the problem of binary linear classification. 

Assume one of $n$ parties is called $P_i$ and has a local data set consisting of instances $x^{(i)}=(x_1^{(i)}, x_2^{(i)},...., x_m^{(i)})$, where $m$ is the number of features, and their corresponding labels $y^{(i)}$.

Party $P_i$ uses its training examples $(x^{(i)},y^{(i)})$ to learn a logistic classifier with weights $w_{i}$. The weights are obtained by solving the following optimization problem:
\begin{align}
     w_{i}= \underset{w}{\arg\min}\frac{1}{t_i}\sum_{k=1}^{t_i} log (1+e^{-y_k^{(i)}f(x_k^{(i)})}),
\end{align}
where $f(x_k^{(i)})=w^Tx_k^{(i)}$ and $t_i$ is the number of training examples of $P_i$.

In order to minimize the loss function, we make use of gradient descent, an iterative optimization algorithm, calculating the optimal $w$ iteratively as 
$w^{j+1} \leftarrow w^j - \alpha \nabla L(w^j)$, where $\alpha$ is the learning rate, $j$ is the iteration, $w^0=0$, and $\nabla L$ is the gradient of the loss function.  Our local logistic regression is a vector-based re-implementation of Jayaraman et al.  \cite{jayaraman2018distributed}

\subsection{Differentially Private Federated Logistic Regression using Output Perturbation by adding Laplace noise}

Privacy-preserving federated learning allows a large number of parties to learn a model while keeping their local training data private.  Parties first train local models on their local data and coordinate with a server to obtain a global model.  Given $n$ parties, let  ${w_i}$, for $i \in 1$ to $n$, represent the local model estimator after minimizing the objective function. 

Then ${W}=\frac{1}{n}\sum_{i=1}^{n}{w_i} + \eta$, where $\eta$ is the differentially private noise added to the cumulative model.

\mathchardef\mhyphen="2D
According to Jayaraman et al \cite{jayaraman2018distributed}, for $1$-$Lipschitz$ the global sensitivity for a multi-party setting is $\frac{2}{n*k*\alpha}$, where k is the size of the smallest dataset amongst the $n$ parties, and $\alpha$ is the regularization parameter. Hence, $\eta=Laplace(\frac{2}{n*k*\alpha*\epsilon})$, where $\epsilon$ is the privacy loss parameter.

In our protocol, each client will add noise to the weights of the trained local model.

\section{Approach}

We illustrate the application of federated learning with differential privacy and secure multi-party computation to a problem of collective interest in finance, that of accurately identifying fraudulent credit card transactions.  This application typifies the case where multiple firms would individually and collectively profit from working together to eliminate the common problem of fraudulent purchases, as the occurrence of fraud benefits none of the lawful parties in the processing chain.

The current limitation to this type of cooperation is data sharing.  The involved companies would not wish to share their local training data, that is their entire history of fraudulent and non-fraudulent transactions, including potentially sensitive customer and merchant information, and in many cases would be legally prohibited from doing so.  A secure federated learning protocol could satisfy the firms and their regulators that data exposure risks have been sufficiently minimized to permit this mutually beneficial collaboration.  Here we describe the key aspects of our approach.

As described in the Introduction, federated learning is an iterative algorithm that follows a simple, repetitive process.  The server chooses some users to produce an updated model.  Those users train a model on their individual data, then send the model updates to the server.  The server aggregates the updates to construct a new global model and shares it with all users.

In this paper, we consider logistic regression as the local learning method, and each client update includes the weights of that logistic regression.  The server receives the weights from all clients at each iteration and computes the new global model using the average of the client updates for each weight.  Recall also from the Introduction the literature demonstrating that the server can infer some private client data from the trained model weights, which is clearly undesirable.

\begin{figure}[h]
\begin{tikzpicture}[scale=0.60]

    \node (a1) at (-5.2,0) {$\boldsymbol{P_1(w_1)}$ };

    \node (a2) at (4.4,0) {$\boldsymbol{P_2(w_2)}$ };

    \node (a3) at (0,-3.3) {$\boldsymbol{P_3(w_3)}$ };
 
    \node (a) at (0,-6.5) {$\boldsymbol{A}$ };
    \node (aTot) at (0,-7.5) {$(\bar{w}_1+\bar{w}_2+\bar{w}_3)/3$ };
    
    \node (a1a2) at (-4.4,0.15) {};
    \node (a1a2') at (3.7,0.15) {};

    \node (a2a1) at (3.7,-0.15) {};
    \node (a2a1') at (-4.4,-0.15) {};
    \node (a21) at (0,0.5) {$r_{12}=r_{21}$};

    \node (a1a3) at (-4.5,-0.3) {};
    \node (a1a3') at (-0.3,-3) {};

    \node (a3a1) at (-0.6,-3.2) {};
    \node (a3a1') at (-4.9,-0.5) {};
    \node (a31) at (-2,-1) {$r_{13}=r_{31}$};
    
    \node (a2a3) at (3.8,-0.3) {};
    \node (a2a3') at (0.3,-3) {};

    \node (a3a2) at (0.6,-3.2) {};
    \node (a3a2') at (4.2,-0.4) {};
    \node (a32) at (1.5,-1) {$r_{32}=r_{23}$};
    
    \node (a1a) at (-5.2,-0.4) {};
    \node (a1a') at (-0.4,-6.4) {};
    \node (a1aVal) at (-2,-4.9) {$\bar{w}_1$};
    
    \node (a2a) at (0,-3.8) {};
    \node (a2a') at (0,-6) {};
    \node (a2aVal) at (0.4,-4.7) {$\bar{w}_3$};
    
    \node (a3a) at (4.6,-0.4) {};
    \node (a3a') at (0.4,-6.4) {};
    \node (a3aVal) at (2,-4.9) {$\bar{w}_2$};

    \draw[<->] (a2a1) to (a2a1');
    \draw[<->] (a1a3) to (a1a3');

    \draw[<->] (a2a3) to (a2a3');

    \draw[->] (a1a) to (a1a');
    \draw[->] (a2a) to (a2a');
    \draw[->] (a3a) to (a3a');
    
\end{tikzpicture}
\caption{ Secure $3$-party weighted average protocol where $\bar{w}_1=w_1+ r_{12}+r_{13},\bar{w}_2=w_2- r_{21}+r_{23},\bar{w}_3=w_3- r_{31}-r_{32}$.} \label{fig:MPC-3}
\end{figure}
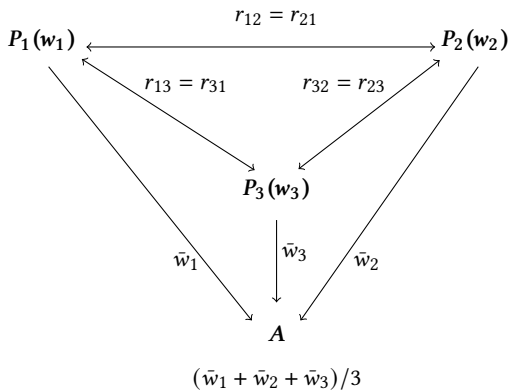

\subsection{Eliminating weight leakage} In order to hide each client's model weights from the server, we use the technique of secure multi-party computation (MPC), in which the clients work together to send their individual updates to the server in an encrypted manner.  In particular, our secure weighted average protocol running across $n$ clients is based on the protocol of Bonawitz et al \cite{Antonio}.

Informally, this can be thought of as each pair of clients sharing a common source of randomness known only to that pair.  When communicating weights to the server, one client within each pair will add, and the other will subtract, the (same) next value in the shared randomness.  In this way, the averaged model produced by the server will be identical to a model produced without MPC, but each weight arriving from each client has $n-1$ large random numbers added to or subtracted from it, removing the server's ability to accurately reconstruct any client's actual model weights.  We describe the process formally in Section \ref{sec:secure-aggr}.

Our protocol includes the ability to reuse the common randomness for each iteration of logistic regression, so the clients will only require pairwise communication once via the server, at the start of the protocol.  In subsequent iterations, each client only has to communicate with the server.  A $3$-party example is given in Figure \ref{fig:MPC-3}.  Note that the weights are encrypted via the use of the common randomness $r$.  The values $\bar{w}$ reveal nothing about the weights $w$.


\begin{table*}[ht]
\begin{protocol}{Privacy-Preserving Federated Logistic Regression Protocol $\PFL$ for a single iteration\\}
\\The protocol $\PFL$ runs with parties $P_1,\ldots, P_n$ and a server $S$. It proceeds as follows:

\textbf{Inputs:} For $i \in [n]$, party~$P_i$ holds input dataset~$D_i$. \\
\textbf{Public Parameters:} $(\mathbb{G},g,q)$ generated by ${\mathcal{G}}(1^{\secparam})$ and modulo $p$. 

{\bf \boldmath $\PFL.\Setup(1^\secparam)$}:

\begin{enumerate}

\item[]{\bf \boldmath Round 1:} Each party $P_i$ for $i\in[n]$ proceeds as follows:
\begin{itemize}
\item  Choose $n$ secrets $a_{i,1},\ldots,a_{i,n}$ uniformly and independently at random from $\mathbb{Z}_q$ and compute $(pk_{i,1},\ldots,pk_{i,n}) = (g^{a_{i,1}}\bmod p,\ldots,g^{a_{i,n}}\bmod p)$. 
\item Generate a Laplace random variable $\eta_{i}$ from $Laplace(\frac{2}{n*len(DS_i)*\alpha*\epsilon})$.

\item Each party $P_i$ sends $pk_{i,j}$ to the Server who forwards $pk_{i,j}$ to party $P_j$. 

\end{itemize}

\item[] {\bf \boldmath Round 2:} Each party $P_j$ for $j\in[n]$ proceeds as follows:
\begin{itemize}
\item Upon receiving all values  $(pk_{1,j},\ldots,pk_{n,j})$, compute the shared common keys $r_{i,j}$ for all $i\in[n]$ as follows:
\begin{enumerate}
    \item Using the secret $a_{j,i}$ compute $c_{i,j}= c_{j,i}=  \left(pk_{i,j}\right)^{a_{j,i}}= \left(g^{a_{i,j}}\right)^{a_{j,i}}\bmod p$.\smallskip
    \item Let $c_{1,j},\ldots,c_{n,j}$ be the set of all common keys. Use a key-derivation function and set $r_{i,j} =  r_{j,i}= H(c_{i,j})$
\end{enumerate}
\end{itemize}

\end{enumerate}

Given the above setup, we can compute the federated logistic regression model as follows: 

\smallskip
{\bf \boldmath $\PFL.{\sf WeightedAverage}(D_i,\{r_{i,j}\}_{j\in[n]})$}:

\begin{enumerate}
\item []\textbf{Round 1:} Each party $P_i$ proceeds as follows: 
\begin{itemize}
    \item Compute the weights $W_{i}$, using Equation (1), of the local logistic classifier obtained by implementing regularized logistic regression on input $D_i$.

The next steps are repeated per weight. Without loss of generality we describe the algorithm for a single weight, denoted by $w_i$. 

\item Compute and send $y_i $ to the server. 
$$y_i := w_i + \sum_{j=i+1}^n r_{i,j} - \sum_{k=1}^{i-1}r_{k,i} + \eta_{i}\bmod p\ .$$

\end{itemize}

\item []\textbf{Round 2:} The server computes $W=\big(\sum_{i=1}^{n}y_i\bmod p \big)/n  $ and sends W to all parties. 
\end{enumerate}

{\bf \boldmath $\PFL.Output(1^\secparam,W)$}: Each party $P_i$ upon receiving $W$ runs the next iteration of the logistic regression repeating $\PFL.WeightedAverage(1^\secparam)$. 

\end{protocol}
\end{table*}

\subsection{Eliminating weighted average leakage} Using our protocol, all information about every client's weights is completely hidden from the server.  However, the shared learned model can still expose some information about individual client weights and subsequently a client's local data set.

Imagine a scenario in which $n-1$ out of the $n$ clients collude or the server chooses $n-1$ adversarial parties. The shared computation result is simply the per-weight average of each client's encrypted weights.  The $n-1$ colluding clients of course know their correct individual weights, but they also know \emph{all} of the encryption added to the weights, because each component of the non-colluding client's encryption occurred with a colluding counterparty who knows the values added or subtracted.  The colluding clients, working together, can thus recover the \emph{exact} model weights of the ``honest'' client.

We would like to avoid client data exposure even in the face of such collusion.  To this end, we can apply differential privacy within the MPC protocol.  In addition to the pairwise encryption, each client will also independently generate and add random ``noise'' to each of its model weights.  Now, even if $n-1$ clients collude, they will only be able to recover the differentially private ``noisy'' weights of the honest client, instead of the exact weights.  The privacy loss parameter in differential privacy is typically called $\epsilon$ (epsilon) and is inversely proportional to the amount of noise added.  Thus there is a trade-off: lower values of epsilon (more ``noise'') better prevent inference of private data during a collusion attack, but eventually interfere with accurate learning of the shared model.

\subsection{Secure Weighted Average Protocol}
\label{sec:secure-aggr}
In this section we formally describe our weighted average protocol $\PFL$, depicted in Protocol 1, for secure logistic regression performed by a set of clients $(P_1,\ldots,P_n)$ and a server $S$.  All operations are performed modulo some bound $p$.

During setup, every pair of parties $P_i$ and $P_j$ will share some common randomness $r_{i,j}=r_{j,i}$. In the online weighted average phase, client $P_i$ sends its weights masked with these common random strings, adding all $r_{i,j}$ for $j > i$ and subtracting all $r_{i,k}$ for $k < i$.  That is, $P_i$ sends to server $S$ the following message for its data $x_i$:
\begin{equation}
  \bar{w}_i := (w_i + \sum_{j = i+1}^n r_{ij} - \sum_{k =1}^{i-1}r_{ki}) \bmod p
\end{equation}
To establish common randomness, each pair of client parties run the Diffie-Hellman Key agreement protocol \cite{DH76} communicating via the server. The cryptographic primitives used in Protocol 1 include:
\begin{itemize}
\item An algorithm ${\mathcal{G}}(1^\secparam)$, where $\secparam$ is the security parameter, that outputs a representation of a cyclic group $\mathbb{G}$ of order $q$ (with $\abs{\abs{q}}=\secparam$) for which the discrete logarithm problem is believed to be hard. Recall that a group $\mathbb{G}$ is cyclic if there exists a generator $g$ such that $\{g^0,g^1,\ldots,g^{q-1}\} = \mathbb{G}$. Moreover, the discrete logarithm problem is believed to be hard if for every probabilistic polynomial time adversary $\A$, there exists a negligible function ${\sf negl}(\cdot)$ such that:
$$ \Pr_{x \leftarrow \mathbb{Z}_q}\left[\A(\mathbb{G},g,q,g^{x}) = x\right] = {\sf negl}(\secparam)$$
In other words, it is hard to guess $x$ given $g^{x}$ for particular groups $\mathbb{G}$.
\item A key derivation function $H: \mathbb{G} \rightarrow \bit^{\secparam}$. It is assumed that if $h$ is distributed uniformly in $\mathbb{G}$, then $H(h)$ is distributed uniformly in $\bit^{\secparam}$. 
\item A pseudorandom generator with double expansion, i.e., $G:\bit^{\secparam} \rightarrow \bit^{2\secparam}$. It is assumed that for every distinguisher $D$ there exists a negligible function ${\sf negl}(\cdot)$ such that:
$$ \left|\Pr_{s \leftarrow \bit^{\secparam}}\left[D(G(s))=1\right] - \Pr_{r \leftarrow \bit^{2\secparam}}\left[D(r) = 1\right]\right| = {\sf negl}(\secparam) $$
\end{itemize}

Protocol 1 is described for a single iteration of the logistic regression. To perform the next iteration the algorithm $\PFL.\Setup$ is not repeated. Instead, the parties can use the common keys $r_{i,j}$ to generate different common keys for the next iteration. More specifically, in the first iteration of the logistic regression we use a pseudorandom generator $(r_{i,j}',s) = G(r_{i,j})$ and update the common randomness $r_{i,j} := r_{i,j}'$. For the next iteration, we run $G(s)$ to obtain a new $r_{i,j}'$ and the seed for the next iteration and so on. Thus the parties need to run the exchange once at the onset of the training.  Our secure weighted average protocol is based on the protocol of Bonawitz et al \cite{Antonio} and its security follows in the same way.

\section{Experiments}

In order to evaluate our method, we implement it in ABIDES, an agent-based interactive discrete event simulation framework described by Byrd et al \cite{byrd2020abides} and available as open-source software from \texttt{https://github.com/abides-sim/abides}.  The selected platform was originally deployed for financial market simulation, but at its core provides a framework easily adapted to other domains.  The system operates in a single-threaded manner to permit deterministic re-simulation in the presence of stochastic elements, but simulates the actions of tens of thousands of agents operating in parallel to one another.  The simulation Kernel tracks time in nanoseconds and enforces ``simulation physics'' including agent computation delays and pairwise noisy communication latency among agents.  Computation delays can be pre-configured or updated during execution to the actual time required for each computation.  All inter-agent communication passes through the Kernel in the form of timestamped messages in a priority queue.  The nature of discrete event simulation permits efficient computation of sparse activity patterns at high time resolution.  Many prior works on federated learning calculate the running time of their protocol ignoring the communication time to the server, but we are able to simulate the latency of the distributed clients' communication.

\begin{table}[ht]
\centering
\begin{tabular}{rrr|rrr}
\toprule
& & & \multicolumn{3}{c}{User}  \\
Users & Total & Server & DH Setup & Training & Encrypt \\
\midrule
100 &  4148.9 & 16.897 &  6.721 & 86.152 &  2.231 \\
200 &  6371.7 & 33.573 & 13.388 & 85.836 &  4.429 \\
300 &  9575.1 & 50.893 & 20.156 & 85.469 &  6.558 \\
400 & 12931.0 & 67.682 & 27.018 & 86.207 &  8.745 \\
500 & 17008.6 & 85.432 & 33.944 & 86.305 & 10.874 \\
\bottomrule
\\
\end{tabular}
\caption {Total protocol time, mean server time per iteration, and mean time per user per protocol iteration in milliseconds for 30 protocol iterations, within which each client runs 250 iterations of local regression training.}
\label{tab:table_simulation_time}
\end{table}

\subsection{Experimental Dataset and Method}
To evaluate the performance of the protocol on real-world data, we selected the Kaggle Credit Card Fraud (2013) dataset \cite{dal2015calibrating}, which provides transformed features that represent the first 26 principal components of unknown original features.  Two original features are provided without transformation: the elapsed time from the start time of the dataset and the amount of the transaction.  We used the Amount column without transformation, but excluded the Time column because our learning method does not attempt to identify temporal clusters or patterns.  We also added a constant intercept feature to permit greater flexibility in the regression.  The dataset provides a categorical $y$ variable identifying whether the transaction was judged fraudulent (True) or not (False).  Of the 284,807 records, only 492 (less than 0.2\%) are labelled fraudulent, representing an extremely unbalanced dataset.

We loaded the dataset once at the start of each complete simulation of the protocol and performed a randomized train-test split (75\% vs 25\%).  At each protocol iteration, each client selected 1000 rows of training data at random as its ``local'' data for that iteration.  The holdout test data was the same for all clients, and no client was ever permitted to train on it.  The clients then implemented Protocol 1 as described in the Approach section, attempting to collaboratively learn a credit card fraud detector, despite each individual client having insufficient data (possibly even zero fraudulent transactions) to do so.  Under Protocol 1, the collaboration is performed in such a way as to not reveal any information about a client's data, using differential privacy within a secure multi-party computation.

\subsection{Protocol Timing Results}

The use of simulation to evaluate the protocol allowed us to construct an accurate model of how long it would take to run such a protocol in the real world.  To accomplish this, each simulation client timed each section of its own part of the protocol, capturing the actual time taken to run the Diffie-Hellman setup (once), the encryption and privacy steps (every iteration), and the local model training step (every iteration).  The service agent captured the actual time taken to receive and store each client's encrypted model each iteration and to combine the models once per iteration.  Our simulation also handles variable communication latency, with each pair of agents having a minimum latency plus a cubic ``jitter'' component that is randomly generated per message.

In Table \ref{tab:table_simulation_time} we provide the timing results of the various steps of our protocol on the credit card fraud dataset with all clients in different areas of New York City, as well as the total (simulated) time required for all parties to complete the protocol and produce a final shared model.  Experiments were run for 30 iterations of Protocol 1, within each of which each client runs 250 iterations of local regression training.

Each experiment was implemented in a single thread on a 24-core Intel Xeon X5650 at 2.6GHZ with 128GB RAM.  Note that while the simulation is single-threaded, it does track a separate current time for each agent, uses the agent's current time when sending or receiving messages, and ensures that agents are not permitted to time travel or perform multiple activities in an overlapping manner.  We assert that these times should therefore be a reasonable estimation of what would occur in a complete, distributed implementation of the protocol.  The time required to perform each simulation, which is \emph{not} the estimated real-world protocol time, ranged from 5 minutes for 100 parties to 28 minutes for 500 parties.  We were able to run many simultaneous simulations (one per core) on the same system with only a slight increase to overall simulation time.

\begin{figure}[ht]
\includegraphics[width=3in]{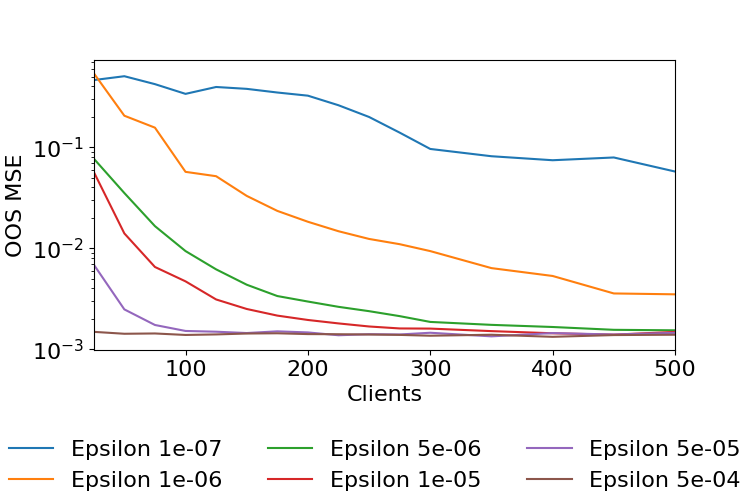}
\caption {Out of sample loss function by number of parties for different values of $\epsilon$ privacy loss parameter.} 
\label{fig:training_loss}
\end{figure}

\begin{figure}[ht]
\includegraphics[width=3in]{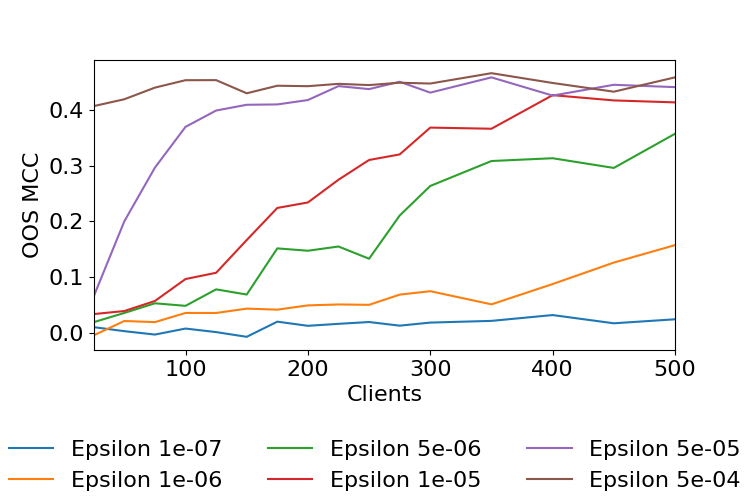}
\caption {Out of sample Matthews Correlation Coefficient by number of parties for different values of $\epsilon$ privacy loss parameter.} 
\label{fig:accuracy_by_parties}
\end{figure}

\subsection{Protocol Accuracy Results}

A key concern with the demonstrated protocol is that, while the secure multi-party computation (MPC) component introduces no accuracy loss to the collaborative training effort, the differential privacy component does in inverse proportion to the $\epsilon$ privacy loss parameter.  In Figure \ref{fig:training_loss}, we show the loss function value on the holdout test data using the securely-learned shared model at the end of the final protocol iteration.  Smaller selections of the privacy loss parameter $\epsilon$ result in more uncertainty about a client's local weights when other parties collude to reveal them, but permit better loss minimization for the same number of parties and protocol iterations.  For instance with 200 clients the training loss and model accuracy worsen dramatically once $\epsilon < 5e^{-5}$.

{\bf Matthews Correlation Coefficient:} Because of the extreme class imbalance in the credit card fraud dataset, simple predictive accuracy would be a poor choice of measure.  The proper classification for 99.8\% of the examples is False (not fraudulent), therefore a naive classifier that always returns False would achieve a misleading 99.8\% accuracy of prediction.

We instead assess our approach using the Matthews Correlation Coefficient (MCC).\cite{matthews1975comparison}  MCC assesses binary classification performance even in the face of unbalanced output classes by accounting for the size of the true negative prediction set: Information not captured by precision, recall, and the F-score.\cite{baldi2000assessing}  MCC is a contingency method of calculating the Pearson product-moment correlation coefficient and therefore has the same interpretation. \cite{pearson1895note,powers2011evaluation,evans1996straightforward}  For example, the MCC of the aforementioned naive classifier would be zero, indicating no correlation between the predicted and actual values.

Let $C(M,D)$ represent the confusion matrix between binary classification model $M$ and data $D$, and recall that for classification output variable $y$, $True$ indicates fraud and $False$ indicates a non-fraudulent transaction.  We can then define MCC for this problem as:
\begin{equation}
    MCC(M,D) = \frac{TP \times TN - FP \times FN}{\sqrt{(TP+FN)(TP+FP)(TN+FP)(TN+FN)}}
\end{equation}
where matrix entries $TP=C(True,True)$, $FP=C(True,False)$, $TN=C(False,False)$, and $FN=C(False,True)$.

In Figure \ref{fig:accuracy_by_parties} we show the MCC of our method's predictions with the correct values.  Smaller selections of privacy loss parameter $\epsilon$ are better for privacy, but directly harm the accuracy of the learned model, so we cannot simply improve privacy by making $\epsilon$ arbitrarily small!  We note that having more parties participate in the computation \emph{does} permit lower values of $\epsilon$ while still producing an accurate shared model.  For all evaluated client population sizes, models trained under Protocol 1 with $\epsilon \geq 5e^{-4}$ had similar mean accuracy to models trained with unsecured (``in the clear'') federated learning.

\subsection{Adversarial Data Recovery}

We now consider that the untrusted server, or a group of other clients, may attempt to recover the unencrypted model weights of an honest client party.  These threats, and the protocol's resistance to them, were discussed in Sections 2 and 3.

\subsubsection*{Snooping Server} First suppose that the server, acting alone, attempts to infer the unencrypted weights of a particular client.  For a single model weight, the value transmitted by the honest party $h$ to the server is:
\begin{equation}
    T = W_h + P_h + R_h
\end{equation}
where $W_h$ is the honest party's original weight, $P_h$ is the differentially private noise added by the honest party, and $R_h=\sum_{c\in C}\pm R_{hc}$ for the set of all other clients $C$.  Recall that for each client pair $(h,c)$, one of them will add $R_{hc}$ and the other will subtract it.

The server has a real problem!  It does not know \emph{any} of the pairwise client values that compose $R_h$.  With 100 participating clients, a single client's $R_h$ is a summation of 99 values randomly generated from the range $(0,2^{32})$.  The value of $R_h$ is thus at least eight orders of magnitude greater than the range of $W_h$, leaving the server with no information about the private weights at all.

\subsubsection*{Colluding Clients} Now suppose that all the clients except one collude to recover the single honest client's locally trained model weights, so they might then make some inference about that client's private data.  Let the honest client be $h$ and the set of $n-1$ colluding clients be $C$.  For a single model weight, the final value contained in the shared model will be a known multiple of:
\begin{equation}
    F = W_h + W_C + P_h + P_C + R - R
\end{equation}
where $W_h$ is the honest party's original weight, $P_h$ is the honest party's differentially private noise, $W_C=\sum_{c\in C}W_c$, $P_C=\sum_{c\in C}P_c$, and $R$ is the sum of all secure multi-party computation (MPC) values.  Note that $R$ is automatically removed when the computation is performed, because for each client pair $i$ and $j$, one client added $R_{ij}$ and the other subtracted $R_{ij}$ from its transmitted values.

The conspirators precisely know terms $W_C$ and $P_C$ and can therefore recover the honest party's $W_h + P_h$.  They cannot accurately infer the privacy noise $P_h$ added by the honest party.  For example the smallest privacy loss parameter value that does not prevent 100 clients from learning the shared model effectively is around $\epsilon = 5e^{-5}$.  Using this, the mean absolute values for the first weight are $W_0=0.62$ and $P_0=0.38$.  Thus there is still considerable uncertainty around the exact weight values of the honest party.

\begin{figure}[ht]
\includegraphics[width=3in]{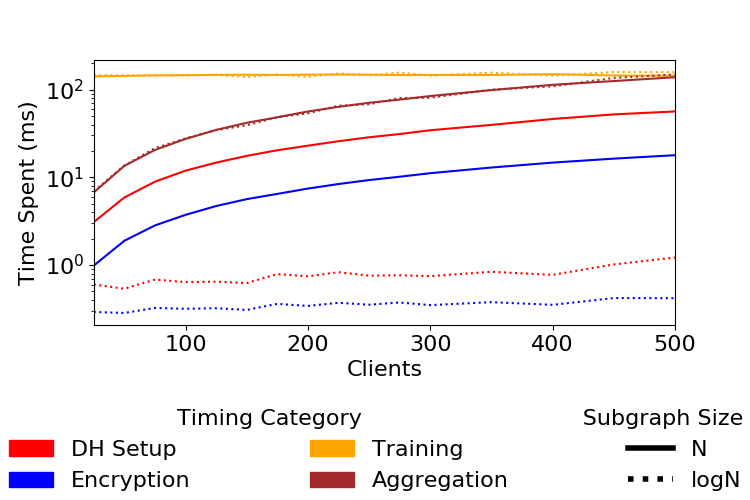}
\caption {Protocol execution time by number of parties, contrasting two MPC peer group sizes.}
\label{fig:logn_timing}
\end{figure}

\begin{figure}[ht]
\includegraphics[width=3in]{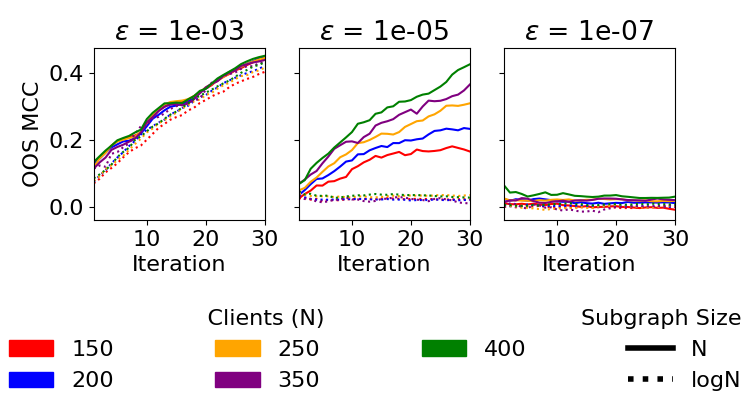}
\caption {Out of sample Matthews Correlation Coefficient by protocol iteration for three values of $\epsilon$ privacy loss parameter, contrasting two MPC peer group sizes.}
\label{fig:logn_accuracy}
\end{figure}

\subsection{Peer Exchange Neighborhood}

We have heretofore assumed that all $N$ clients will partner with all other clients in the secure multi-party computation (MPC) peer exchange.  That is, each client will modify their differentially private weights with $N-1$ random large numbers, added or subtracted.  This is ideal from a privacy perspective, because it means $N-1$ clients must collude to accurately remove the MPC encryption from one client's transmitted weights.  However, for large client populations $N$, it imposes a growth factor of $O(N^2)$ to the overall protocol effort for these pairwise exchanges.  It may therefore be acceptable to allow smaller ``neighborhoods'' (subgraphs) of MPC peer exchange.  In Figures \ref{fig:logn_timing} and \ref{fig:logn_accuracy}, we provide the mean execution time by task category, and the shared model accuracy, for peer exchange subgraphs of size $N$ and $\log(N)$.  Depending on the selection of privacy loss parameter $\epsilon$, the use of a smaller peer group for key exchange provides a significant speed boost to certain elements of the protocol, but at the expense of either privacy or accuracy.

\section{Conclusion}
We have presented a differentially private secure multi-party computation protocol for federated learning intended to be accessible to an audience with some computation background, but without requiring prior knowledge of security, encryption, privacy, or distributed learning.

We have demonstrated the techniques on a common financial data set containing two days of anonymized credit card transactions, of which about 0.2\% are labeled as fraudulent, and shown how these techniques permit multiple parties (e.g. financial firms) to collaboratively learn a useful fraud detection model \emph{without} sharing any of their client data or transactions, and with added layers of protection that make private data recovery difficult or impossible even with adversarial actors participating in the system.

Finally, we have implemented the full protocol on a real-world financial data set in an agent-based interactive discrete event simulation and conducted experiments to evaluate the accuracy and expected running time of the protocol for various numbers of participating parties, various values of the $\epsilon$ privacy loss parameter, and a multi-party computation neighborhood size of $N$ vs $\log(N)$ peers.  

\begin{acks}
This material is based upon research supported by the National Science Foundation under Grant No. 1741026 and by a JP Morgan PhD Fellowship.

This paper was prepared for informational purposes in part by the Artificial Intelligence Research group of JPMorgan Chase \& Co. and its affiliates (“J.P. Morgan”), and is not a product of the Research Department of J.P. Morgan.  J.P. Morgan makes no representation and warranty whatsoever and disclaims all liability, for the completeness, accuracy or reliability of the information contained herein.  This document is not intended as investment research or investment advice, or a recommendation, offer or solicitation for the purchase or sale of any security, financial instrument, financial product or service, or to be used in any way for evaluating the merits of participating in any transaction, and shall not constitute a solicitation under any jurisdiction or to any person, if such solicitation under such jurisdiction or to such person would be unlawful.
\end{acks}

\bibliographystyle{ACM-Reference-Format}
\bibliography{smpai}

\end{document}